\documentclass[twocolumn,showpacs,preprintnumbers,amsmath,amssymb]{revtex4}
\usepackage{graphicx}
\usepackage{dcolumn}
\usepackage{bm}

\begin{document}
%


\title{Bell's Theorem Without Inequalities for two Maximally Entangled Particles}
\author{W. LiMing}
\email{wliming@scnu.edu.cn}
\author{Z. L. Tang}
\affiliation{Dept. of Physics, South-China Normal University, Guangzhou
510631, P. R. China}
\author{C. Liao}
\affiliation{School for Information and Optoelectronic Science and Engineering,
South-China Normal University, Guangzhou 510631, P. R. China}
\date{\today}

\begin{abstract}
A proof of Bell's theorem without inequalities for two maximally entangled
particles is proposed using the technique of quantum teleportation. It follows
Hardy's arguments for a non-maximally entangled state with the help of two auxiliary
particles without correlation. The present proof can be tested by
measurements with 100\% probability.
\end{abstract}

\maketitle
Bell's theorem claims that quantum mechanics cannot be reproduced with the hidden
variable local model\cite{Bell}.
Recently, Bell's theorem
appeared in forms without inequalities\cite{GHZ, Hardy, Adan1, Adan2}, exhibiting greater
contradiction between the local model and quantum mechanics.
Greenberger et al (GHZ) proposed a proof for three entangled particles, thus three
observers are needed\cite{GHZ}.
Hardy proposed a proof for  two non-maximally entangled particles\cite{Hardy}.
Cabello proposed a GHZ-like and more delicate proof for two pairs of maximally entangled
particles\cite{Adan1, Adan2}. It seems that non-locality exists only in
a maximally entangled state of three or more particles or in a non-maximally entangled
state of two
particles.  The paradox of Einstein, Podolsky, and Rosen (EPR), however,
argues the non-locality
of two maximally entangled particles\cite{EPR}.
The question is whether it is possible to demonstrate Bell's theorem without using
inequalities by two maximally entangled particles.

In this paper I propose a proof of Bell's theorem without using inequalities for two
maximally entangled particles, using the technique of quantum teleportation. It follows
Hardy's arguments for the contradiction between the hidden variable local model and the
non-locality of quantum mechanics.

Consider two spin-$1/2$ particles, 1 and 2, in a maximally entangled state, i.e.,
a spin singlet:
\begin{equation}
\label{entangled} |\Psi\rangle_{12} = \frac{1}{\sqrt{2}}
(|+\rangle_1|-\rangle_2 - |-\rangle_1 |+\rangle_2)
\end{equation}
where "+" and "-" denote, respectively, spin up and down.
These two particles are transmitted in opposite directions to two observers, Alice
and Bob, with a space-like separation.
Both Alice and Bob have prepared an auxiliary particle, A and B, respectively,
with the following spin states
\begin{equation}
|A\rangle =|+\rangle \,\,\,\,\,,\,\,\,\, |B\rangle
=(|+\rangle+|-\rangle)/\sqrt{2},
\end{equation}
which are the eigenstates of $\sigma_z$ and $\sigma_x$, respectively.
These two auxiliary particles are not correlating with each other and never
transmitted between the two observers, thus are considered as part of
the apparatus. The total
spin state of this four-particle system is given by
\begin{equation}
|\Psi\rangle = |A\rangle |\Psi\rangle_{12} |B\rangle
\end{equation}
In order to use the technique of quantum teleportation this state is expanded
in the Bell basis
of Alice's particles A and 1 as follows:
\begin{eqnarray}
|\Psi\rangle=-\frac{1}{2}&\bigl[&|\Psi^-\rangle_{A1}|+\rangle_2 \nonumber \\
&+&|\Psi^+\rangle_{A1}|+\rangle_2 \nonumber \\
&-&|\Phi^-\rangle_{A1} |-\rangle_2 \nonumber \\
&-&|\Phi^+\rangle_{A1}) |-\rangle_2\bigr] |B\rangle \label{12}
\end{eqnarray}
where the Bell states are given by
\begin{eqnarray}
|\Psi^\pm\rangle&=&\frac{1}{\sqrt{2}} (|+\rangle |-\rangle \pm |-\rangle |+\rangle) \\
|\Phi^\pm\rangle&=&\frac{1}{\sqrt{2}} (|+\rangle |+\rangle \pm
|-\rangle |-\rangle)
\end{eqnarray}
$|\Psi\rangle$ can also be expanded in the Bell basis of Bob's
particles 2 and B as follows:
\begin{eqnarray}
|\Psi\rangle=-\frac{1}{2\sqrt{2}}|A\rangle
 \bigl[(|+\rangle_1&+&|-\rangle_1)|\Psi^-\rangle_{2B} \nonumber \\
+(|+\rangle_1&-&|-\rangle_1)|\Psi^+\rangle_{2B}   \nonumber \\
             -(|+\rangle_1&+&|-\rangle_1)|\Phi^-\rangle_{2B} \nonumber \\
- (-|+\rangle_1&+&|-\rangle_1)|\Phi^+\rangle_{2B}\bigr] \label{34}
\end{eqnarray}

According to the principle of quantum teleportation, if Alice
measures one of the Bell states of her two particles, A and 1,
particle 2 will collapse to the corresponding quantum state, see
(\ref{12}). For example, if Alice measures the Bell state,
$|\Psi^-\rangle_{A1}$, particle 2 collapses to state $|+\rangle$.
In the same way, if Bob measures the state, $|\Psi^-\rangle_{2B}$,
particle 1 collapses to state $(|+\rangle+|-\rangle)/\sqrt{2}$,
see (\ref{34}). It is seen that the state of particle A of Alice
is copied into the state of particle 2 of Bob without transmitting
any physical information between Alice and Bob\cite{Bouwnmeester,
Bennet}. The above two measurements correspond to the following
two projecting operators:
\begin{eqnarray}
\hat D_1&=&|\Psi^-\rangle_{A1}\langle\Psi^-| \nonumber \\
\hat D_2&=&|\Psi^-\rangle_{2B}\langle\Psi^-|
\end{eqnarray}

Including another two operators $\hat U_1$ and $\hat U_2$,
\begin{equation}
\hat U_1= |+\rangle_1\langle+| \,\,\,\,,\,\,\,\,\, \hat U_2 =
|+\rangle_2\langle+|
\end{equation}
one has four physical observable quantities, $\hat D_1,\hat D_2, \hat U_1, \hat U_2$.
They take values 0 or 1 corresponding to their eigenvalues, denoted as
$D_1,D_2, U_1, U_2$.

Now if Alice measures $\hat U_1$ on particle 1, and Bob measures
$\hat U_2$ on particle 2, they have
\begin{equation}\label{u1u2}
U_1 U_2=0 .
\end{equation}
This is because, since particles 1 and 2 are in a maximally entangled state,
(\ref{entangled}), their spins are always opposite to each other.

From (\ref{12}), if Alice measures $\hat D_1$ on her particles A and 1,
and Bob measures $\hat U_2$ on his particle 2, they have
\begin{equation}\label{d1u2}
\text{if} \qquad  D_1=1 \qquad \text{then}\qquad U_2=1
\end{equation}
From (\ref{34}), if Bob measures $\hat D_2$ on his particles 2 and B, and
Alice measures $\hat U_1$ on his particle 1, they have
\begin{equation}\label{d2u1}
\text{if} \qquad  D_2=1 \qquad \text{then}\qquad U_1=1
\end{equation}

The most important fact here is $[\hat D_1, \hat D_2] =0$, indicating that
$\hat D_1$ and $\hat D_2$ can be measured simultaneously. This allows Alice
and Bob to make a joint measurement for these two observable quantities on the system.
The probability of the system with $D_1=D_2=1$  is given by
\begin{equation}
P=\langle\Psi|\hat D_1 \hat D_2|\Psi\rangle = \frac{1}{16}\qquad
\text{for}\,\,\, D_1=D_2=1 \label{prob}
\end{equation}

Now Hardy's arguments can be applied to deduce a contradiction
between the non-locality of the entangled state, $|\Psi\rangle$,
and the hidden variable local model\cite{Hardy}. The latter claims
that Alice's choice of measurement cannot influence the outcome of
any measurement of Bob, since there is a space-like space-time
separation between Alice and Bob. For example, for a run that
Alice and Bob obtain $D_1=U_2=1$, if Alice had measured something
else, say $\hat U_1$ for particle 1, instead of $\hat D_1$, he
would not affect the outcome of Bob's measurement $U_2=1$, {\it
vice versa}. In another word, due to (\ref{d1u2}), if Alice
obtained $D_1=1$, she can predict without any uncertainty the
outcome of Bob's measurement, $U_2=1$. According to EPR's
argument, $U_2=1$ is a physical reality element of particle 2.

Consider a run of a joint measurement that Alice and Bob found $D_1=D_2=1$.
This run does exist since the probability
exists, see (\ref{prob}). In the hidden variable local model, we have the
following deductions:

{\it Deduction 1}: From the fact that we have $D_1=1$ it follows from
(\ref{d1u2}) that if $\hat U_2$ had been measured we would obtain $U_2=1$.
In another word,  $U_2=1$ is a
physical reality element of particle 2.

{\it Deduction 2}: Since Alice's choice of measurement does not affect the
outcome of Bob's measurement, even if $\hat U_1$ had been measured on particle 1
instead of $\hat D_1$ we would still have $U_2=1$.

{\it Deduction 3}: By a similar argument we can deduce from the fact $D_2=1$
and (\ref{d2u1}) that $U_1=1$. In another word, $U_1=1$ is a physical reality
element of particle 1.

{\it Deduction 4}: Thus, for this run, we have $U_1U_2=1$. Hence, if we had
measured $\hat U_1$ and $\hat U_2$ instead of  $\hat D_1$ and $\hat D_2$, we
would have obtained   $U_1U_2=1$, which, however, contradicts (\ref{u1u2}).

It is seen that from the hidden variable local model we arrive at a
contradiction and therefore, an entangled state of quantum mechanics must be
nonlocal.  Physical reality element does not exist.

To demonstrate this proof experimentally, one needs first to verify (\ref{u1u2},
\ref{d1u2},\ref{d2u1}), and then measure   $\hat D_1$ and $\hat D_2$. If
$D_1=D_2=1$ is observed then non-locality of quantum mechanics is verified.

In the above formalism, only one pair of Bell states, i.e.,
$|\Psi^-\rangle_{A1}$ and $|\Psi^-\rangle_{2B}$, are considered.
In fact, operators $\hat D_1$ and $\hat D_2$
can be made up of each pair of Bell states, such as \\
$\hat D_1=|\Psi^-\rangle_{A1}\langle\Psi^-|$ and
$\hat D_2=|\Psi^+\rangle_{2B}\langle\Psi^+|$,......\\
$\hat D_1=|\Psi^+\rangle_{A1}\langle\Psi^+|$ and
$\hat D_2=|\Phi^-\rangle_{2B}\langle\Phi^-|$,......\\

Totally, one has $4 \times 4=16$ different pairs of
$\hat D_1$ and $\hat D_2$. This is obviously true because the four Bell states compose
a complete basis. For each pair of operators $\hat D_1$ and $\hat D_2$ one
defines corresponding
operators $\hat U_1$ and $\hat U_2$. Using each group of operators
$\hat D_1,\,\,\hat D_2,\,\,\hat U_1,\,\,\hat U_2$
one can deduce the above nonlocal property of entangled states.
Therefore, from (\ref{prob}), the total probability to
test the Bell theorem is $16 \times 1/16 = 100\%$.

In summary, Bell's theorem without inequalities is proved
for a maximally entangled state based on the technique of quantum teleportation.
Auxiliary particles A and B have not been transmitted
between Alice and Bob and are not correlating to each other, thus are taken as
part of the apparatus. The non-locality proved here belongs to the maximally
entangled state of particles 1 and 2. This proof can be tested
by measurements with 100\% probability.

\begin{acknowledgments}
This work was supported by the National Fundamental Research Program under Grant No
2001CD309300.
\end{acknowledgments}


\end{document}